## PLANETARY SCIENCE

# The Pluto system: Initial results from its exploration by New Horizons

S. A. Stern,[1]* F. Bagenal,[2] K. Ennico,[3] G. R. Gladstone,[4] W. M. Grundy,[5]
W. B. McKinnon,[6] J. M. Moore,[3] C. B. Olkin,[1] J. R. Spencer,[1] H. A. Weaver,[7]
L. A. Young,[1] T. Andert,[8] J. Andrews,[1] M. Banks,[9] B. Bauer,[7] J. Bauman,[10]
O. S. Barnouin,[7] P. Bedini,[7] K. Beisser,[7] R. A. Beyer,[3] S. Bhaskaran,[11] R. P. Binzel,[12]
E. Birath,[1] M. Bird,[13] D. J. Bogan,[14] A. Bowman,[7] V. J. Bray,[15] M. Brozovic,[11]
C. Bryan,[10] M. R. Buckley,[7] M. W. Buie,[1] B. J. Buratti,[11] S. S. Bushman,[7] A. Calloway,[7]
B. Carcich,[16] A. F. Cheng,[7] S. Conard,[7] C. A. Conrad,[1] J. C. Cook,[1] D. P. Cruikshank,[3]
O. S. Custodio,[7] C. M. Dalle Ore,[3] C. Deboy,[7] Z. J. B. Dischner,[1] P. Dumont,[10]
A. M. Earle,[12] H. A. Elliott,[4] J. Ercol,[7] C. M. Ernst,[7] T. Finley,[1] S. H. Flanigan,[7]
G. Fountain,[7] M. J. Freeze,[7] T. Greathouse,[4] J. L. Green,[17] Y. Guo,[7] M. Hahn,[18]
D. P. Hamilton,[19] S. A. Hamilton,[7] J. Hanley,[1] A. Harch,[20] M. Hart,[7] C. B. Hersman,[7]
A. Hill,[7] M. E. Hill,[7] D. P. Hinson,[21] M. E. Holdridge,[7] M. Horanyi,[2] A. D. Howard,[22]
C. J. A. Howett,[1] C. Jackman,[10] R. A. Jacobson,[11] D. E. Jennings,[23] J. A. Kammer,[1]
H. K. Kang,[7] D. E. Kaufmann,[1] P. Kollmann,[7] S. M. Krimigis,[7] D. Kusnierkiewicz,[7]
T. R. Lauer,[24] J. E. Lee,[25] K. L. Lindstrom,[7] I. R. Linscott,[26] C. M. Lisse,[7]
A. W. Lunsford,[23] V. A. Mallder,[7] N. Martin,[20] D. J. McComas,[4] R. L. McNutt Jr.,[7]
D. Mehoke,[7] T. Mehoke,[7] E. D. Melin,[7] M. Mutcher,[27] D. Nelson,[10] F. Nimmo,[28]
J. I. Nunez,[7] A. Ocampo,[17] W. M. Owen,[11] M. Paetzold,[18] B. Page,[10] A. H. Parker,[1]
J. W. Parker,[1] F. Pelletier,[10] J. Peterson,[1] N. Pinkine,[7] M. Piquette,[2] S. B. Porter,[1]
S. Protopapa,[19] J. Redfern,[1] H. J. Reitsema,[20] D. C. Reuter,[23] J. H. Roberts,[7]
S. J. Robbins,[1] G. Rogers,[7] D. Rose,[1] K. Runyon,[7] K. D. Retherford,[4]
M. G. Ryschkewitsch,[7] P. Schenk,[29] E. Schindhelm,[1] B. Sepan,[7] M. R. Showalter,[21]
K. N. Singer,[1] M. Soluri,[30] D. Stanbridge,[10] A. J. Steffl,[1] D. F. Strobel,[31] T. Stryk,[32]
M. E. Summers,[33] J. R. Szalay,[2] M. Tapley,[4] A. Taylor,[10] H. Taylor,[7] H. B. Throop,[9]
C. C. C. Tsang,[1] G. L. Tyler,[26] O. M. Umurhan,[3] A. J. Verbiscer,[34] M. H. Versteeg,[4]
M. Vincent,[1] R. Webbert,[7] S. Weidner,[7] G. E. Weigle II,[4] O. L. White,[3]
K. Whittenburg,[7] B. G. Williams,[10] K. Williams,[10] S. Williams,[7] W. W. Woods,[26]
A. M. Zangari,[1] E. Zirnstein[4]

The Pluto system was recently explored by NASA's New Horizons spacecraft, making closest approach on 14 July 2015. Pluto's surface displays diverse landforms, terrain ages, albedos, colors, and composition gradients. Evidence is found for a water-ice crust, geologically young surface units, surface ice convection, wind streaks, volatile transport, and glacial flow. Pluto's atmosphere is highly extended, with trace hydrocarbons, a global haze layer, and a surface pressure near 10 microbars. Pluto's diverse surface geology and long-term activity raise fundamental questions about how small planets remain active many billions of years after formation. Pluto's large moon Charon displays tectonics and evidence for a heterogeneous crustal composition; its north pole displays puzzling dark terrain. Small satellites Hydra and Nix have higher albedos than expected.

luto was discovered in 1930 (*1*); it forms a binary planet with its moon Charon, and the system's basic properties have been measured remotely from Earth (*1*). Pluto was long thought to be a misfit or anomaly in the solar system. However, the 1992 discovery of the Kuiper Belt (*2*)—provided important context demonstrating that Pluto is the largest of a class of small planets formed in the outer solar system during the ancient era of planetary accretion ~4.5 billion years ago.

New Horizons (*3*) launched on 19 January 2006 and successfully completed a close approach to the Pluto system on 14 July 2015 at a distance of 13,691 km from Pluto's center. It carries a sophisticated suite of instruments summarized in (*4*), including the Ralph multicolor/panchromatic mapper and mapping infrared (IR) composition spectrometer; the Long Range Reconnaissance Imager (LORRI), a long-focal length panchromatic visible imager; the Alice extreme/far ultraviolet (UV) mapping spectrograph; the REX (Radio Experiment); the Solar Wind Around Pluto (SWAP) instrument; PEPSSI (Pluto Energetic Particle Spectrometer Science Instrument); and the VBSDC (Venetia Burney Student Dust Counter), a dust impact detector.

This article contains the first scientific results and post-flyby interpretations from the New Horizons Pluto flyby, organized according to the objects in the system.

## Pluto
### Geology and imaging

New Horizons has so far provided coverage (Fig. 1A) of the near-encounter, anti–Charon-facing hemisphere north of 30° south latitude at 2.2 km/pixel, with limited areas on that hemisphere covered at a higher resolution of 400 m/pixel. On the Charon-facing opposite hemisphere of Pluto, imaging resolution varies from 13 to 27 km/pixel. Dynamical and physical properties of Pluto and its satellites are given in Table 1. Limb fits using full-disk images, combined in a joint solution, give a mean radius for Pluto of 1187 ± 4 km (*5*), at the larger end of a previously uncertain range of 1150 to 1200 km (*6*). No oblateness has been detected (*5*), yielding a conservative upper limit on Pluto's polar flattening (a difference of <12 km between equatorial and polar axes) of 1%. We conclude from this that Pluto does not record significant shape evidence of an early, high-spin period after Pluto-Charon binary formation (*7*), presumably because it was warm and deformable during or after tidal spindown.

Pluto displays a diverse range of landforms, as well as evidence for geological and other processes that have substantially modified its surface up to geologically recent times. Pluto's latitudinal band from about 25°S to 10°N features large, discrete expanses of low-albedo terrain interspersed with brighter regions. Terrain that is more reflective generally occurs in mid- and high latitudes. The large, prominent high-albedo region of the New Horizons encounter hemisphere that we call Tombaugh Regio (TR; all surface feature names currently used are informal) straddles the equator on the anti-Charon hemisphere (Fig. 2A).

[1]Southwest Research Institute, Boulder, CO 80302, USA. [2]Laboratory for Atmospheric and Space Physics, University of Colorado, Boulder, CO 80303, USA. [3]National Aeronautics and Space Administration (NASA) Ames Research Center, Space Science Division, Moffett Field, CA 94035, USA. [4]Southwest Research Institute, San Antonio, TX 28510, USA. [5]Lowell Observatory, Flagstaff, AZ 86001, USA. [6]Department of Earth and Planetary Sciences, Washington University, St. Louis, MO 63130, USA. [7]Johns Hopkins University Applied Physics Laboratory, Laurel, MD 20723, USA. [8]Universität der Bundeswehr München, Neubiberg 85577, Germany. [9]Planetary Science Institute, Tucson, AZ 85719, USA. [10]KinetX Aerospace, Tempe, AZ 85284, USA. [11]NASA Jet Propulsion Laboratory, La Cañada Flintridge, CA 91011, USA. [12]Massachusetts Institute of Technology, Cambridge, MA 02139, USA. [13]University of Bonn, Bonn D-53113, Germany. [14]NASA Headquarters (retired), Washington, DC 20546, USA. [15]University of Arizona, Tucson, AZ 85721, USA. [16]Cornell University, Ithaca, NY 14853, USA. [17]NASA Headquarters, Washington, DC 20546, USA. [18]Rheinisches Institut für Umweltforschung an der Universität zu Köln, Cologne 50931, Germany. [19]Department of Astronomy, University of Maryland, College Park, MD 20742, USA. [20]Southwest Research Institute, Boulder, CO 80302, USA. [21]Search for Extraterrestrial Intelligence Institute, Mountain View, CA 94043, USA. [22]Department of Environmental Sciences, University of Virginia, Charlottesville, VA 22904, USA. [23]NASA Goddard Space Flight Center, Greenbelt, MD 20771, USA. [24]National Optical Astronomy Observatory, Tucson, AZ 26732, USA. [25]NASA Marshall Space Flight Center, Huntsville, AL 35812, USA. [26]Stanford University, Stanford, CA 94305, USA. [27]Space Telescope Science Institute, Baltimore, MD 21218, USA. [28]University of California, Santa Cruz, CA 95064, USA. [29]Lunar and Planetary Institute, Houston, TX 77058, USA. [30]Michael Soluri Photography, New York, NY 10014, USA. [31]Johns Hopkins University, Baltimore, MD 21218, USA. [32]Roane State Community College, Jamestown, TN 38556, USA. [33]George Mason University, Fairfax, VA 22030, USA. [34]Department of Astronomy, University of Virginia, Charlottesville, VA 22904, USA.
*Corresponding author. E-mail: astern@boulder.swri.edu

**Fig. 1. Pluto surface imaging results.** (**A**) Simple cylindrical mosaic of Pluto; the area covered by the seven highest-resolution (400 m/pixel) frames is highlighted by a red boundary and is shown in (C). (**B**) Polar stereographic mosaic of Pluto's north pole. (**C**) Seven-image, 400 m/pixel mosaic covering the majority of Sputnik Planum (SP), with Norgay Montes and Hillary Montes bordering to the south. Areas shown in (D) and (E) are labeled d and e. (**D**) Detail of the northern margin of SP: (1) Polygonal terrain at the northern margin of SP. (2) Rugged, cratered terrain north of SP. (3) Patterns indicative of viscous flow in the ice (see text). (4) A crater (diameter ~40 km) that has been breached by ice from SP. (**E**) Detail of SP's southwest margin. (5) Polygonal terrain in SP. (6) Ice plains separating Hillary Montes from the dark terrain of Cthulhu Regio (CR). The ice appears to cover and embay portions of CR. (7) Ice infilling a crater at SP's margin. (8) Rough, undulating terrain south of Norgay Montes displaying very few impact craters at 400 m/pixel. (**F**) Histograms of the I/F distributions of Pluto and Charon. Note the wider range of I/F values for Pluto than for Charon.

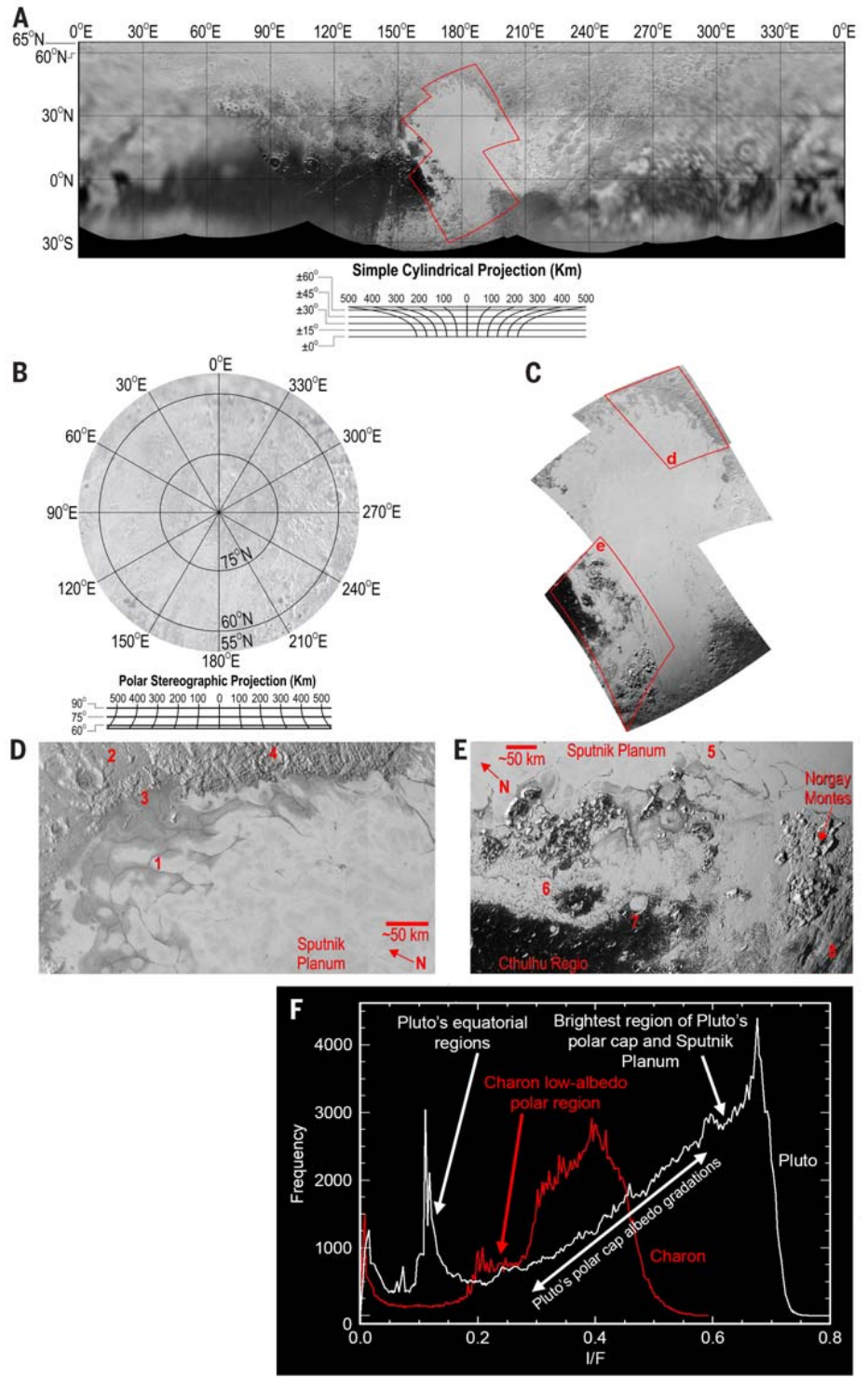

TR measures about 1800 km east to west and 1500 km north to south.

At 2.2 km/pixel, widely distributed impact craters up to 260 km in diameter are seen in the near-encounter hemisphere. Many appear to be substantially degraded or infilled, and some are highlighted by bright ice-rich deposits on their rims and/or floors. This includes the dark equatorial terrain immediately west of TR, called

Cthulhu Regio (CR), which appears densely cratered. Tectonic features, including scarps and troughs up to 600 km in length, occur within and to the north of CR.

A large, apparently level plains unit we call Sputnik Planum (SP) constitutes the west half of TR. Several physiographic provinces have been identified in this region (Figs. 1 and 2). Mountains locally rise 2 to 3 km above their surrounding

**Table 1. Properties of the Pluto-Charon system.** Boldface entries are values from New Horizons. Mean orbital elements (semimajor axis, orbital period, eccentricity, and inclination) for Charon are Plutocentric, whereas those for the small satellites are barycentric and are based on numerical integrations (*35*); *GM* (standard gravitational parameter) values are also from (*35*).

| Body | Semimajor axis (km) | Period (days) | Eccentricity | Inclination (degrees) | Radius (km) | *GM* (km³ s⁻²) | Density (kg m⁻³) |
|------|---------------------|---------------|--------------|-----------------------|-------------|------------------|-------------------|
| Pluto | | 6.3872 | | | **1187 ± 4*** | 869.6 ± 1.8 | **1860 ± 13** |
| Charon | 19,596 | 6.3872 | 0.00005 | 0.0 | **606 ± 3*** | 105.88 ± 1.0 | **1702 ± 21** |
| Styx | 42,413 | 20.1617 | 0.00001 | 0.0 | 1.8 to 9.8† | 0.0000 ± 0.0001 | |
| Nix | 48,690 | 24.8548 | 0.00000 | 0.0 | 54 × 41 × 36‡ | 0.0030 ± 0.0027 | |
| Kerberos | 57,750 | 32.1679 | 0.00000 | 0.4 | 2.6 to 14† | 0.0011 ± 0.0006 | |
| Hydra | 64,721 | 38.2021 | 0.00554 | 0.3 | 43 × 33‡ | 0.0032 ± 0.0028 | |

*From limb fits to LORRI images; radius error is pixel scale of best resolved image for each. Pluto's radius is consistent with radio occultation results as well; see (*36*) for technique.   †From (*32*).   ‡Axial dimensions derived from LORRI and MVIC images (see text).

**Fig. 2. Maps with informal feature names used on Pluto (A) and Charon (B).** Geomorphological regions that we consider to be distinct are colored as follows: red, terrae; green, craters; light yellow, chasmata; orange, maculae; blue, montes; purple, regios; yellow, plana; cyan, dorsae; pink, cavi; light green, lineae; golden yellow, colles; red lines, rupes; green lines, fossae; yellow lines, valles.

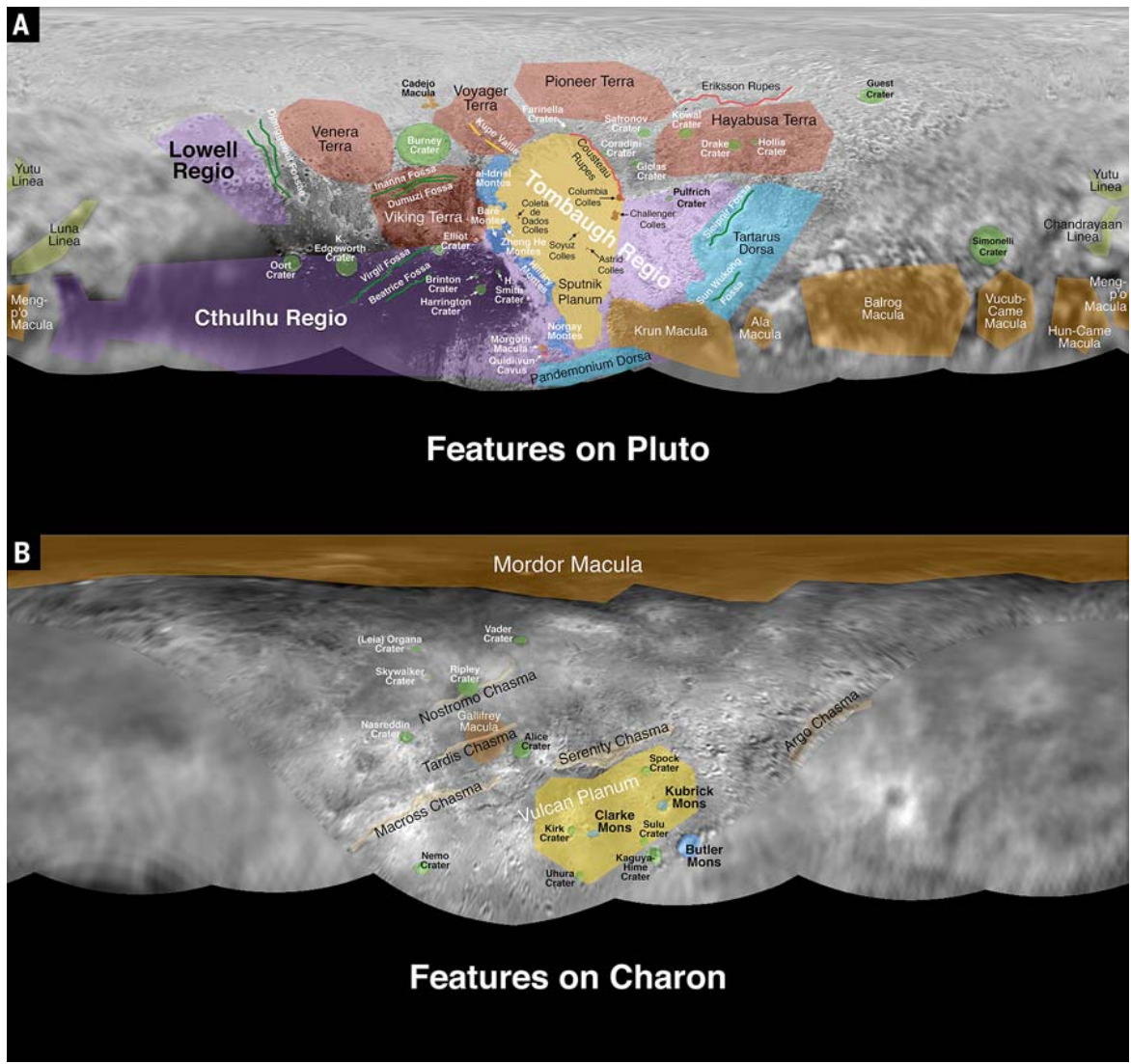

terrain, as calculated from shadow length measurements. These and other high, steep-sided topographic features seen across Pluto require materials that will not relax under their own weight on geologic time scales. The nitrogen (N₂), carbon monoxide (CO), and methane (CH₄) ices that were known from ground-based spectroscopy (*1*) to dominate Pluto's visible surface would collapse exceptionally rapidly (*8, 9*) because they are weak, van der Waals–bonded solids. The mountains detected by New Horizons imagery therefore imply the presence of a widespread, stronger, presumably water ice–based, solid "bedrock." We further conclude that the observed N₂, CO, and CH₄ ices must only be a surface veneer above this bedrock.

Portions of the mountainous terrain are broken into hummocky regions of varying scale. An undulating, lightly cratered terrain occurs at the south end of Fig. 1C and in a large region at the eastern edge of TR; its broadly rounded undulations are separated by linear depressions and troughs. The hummocks range from 20 to 150 km across and a few hundred meters in relief (as derived principally from shadow measurements) and feature smaller superimposed, rounded ridges. This terrain may be tectonic in origin.

SP (Fig. 1, C and D) has no confirmed craters. Much of its surface is divided into polygonal and ovoid-shaped cells tens of kilometers wide, themselves bordered by shallow troughs of characteristic width 2 to 3 km. Some troughs have darker material within them and some are traced by clumps of hills that rise up to a few hundred meters above the surrounding terrain; others contain narrow medial ridges tens of meters high. Around the margins of SP, portions of the surface appear to be etched by fields of small pits that may have formed by sublimation. Aligned dark streaks in SP are tentatively interpreted as wind streaks (fig. S1). The central, brightest region of SP contains $N_2$ and $CH_4$ ices and also coincides with a surface enhancement in CO ice (see below). SP is mostly bordered by locally higher terrain, which suggests that it fills a topographic basin.

Some features of SP suggest bulk flow similar to terrestrial glaciers. Two lobes with sharp margins extend south; topographic shading suggests a convex upward profile (Fig. 1C, bottom). Along the northern margin of SP, hills of apparent basement materials protrude above the smooth terrain (possibly water-ice nunataqs). Albedo features on SP's smooth terrain appear to be diverted around these hills (Fig. 1D), suggesting flow around obstacles. Elsewhere, SP material embays the interior of a degraded crater through a rim breach (Fig. 1D). Such bulk flow driven by modest topographic gradients is consistent with the rheological characteristics of $N_2$, CO, or $CH_4$ ices at Pluto surface conditions (i.e., near Pluto's ~38 K surface temperature) (9).

The origin of the polygonal and ovoid features on SP is uncertain. They could be the surface manifestation of contraction (analogous to mud or cooling cracks), or insolation-related processes, or the result of fracture of the surface due to extension and/or uplift of the subsurface, but they are perhaps most consistent morphologically with solid-state convection [see, e.g., (10)]. Internal convection is also consistent with evidence cited above for the flow of the material that fills TR, in that the surface layer apparently possesses a low enough viscosity that it can creep or flow under low driving stresses.

Varying crater abundances indicate wide-ranging surface ages on Pluto, in the sense that numerous large craters are seen on certain regions (such as CR), whereas no craters with diameters of >10 km can be identified on SP. Model ages for SP derived from estimates of Kuiper Belt bombardment (see the discussion of Charon crater counts below) imply active geomorphic processes within the last few hundred million years (11, 12) and possibly continuing to the present. Such re-surfacing can occur via surficial erosion/deposition (as at Titan), crater relaxation (as at Enceladus), crustal recycling or tectonism (as at Europa), or some combination of these processes (13). For icy satellites, resurfacing is generally associated with eccentricity tides (14), but these are not a viable heat source today for Pluto or Charon, whose orbital eccentricities are fully damped (Table 1); as such, the young surface units on Pluto present a puzzle regarding the energy source(s) that power such resurfacing over time scales of billions of years.

### Surface color and composition

The radiance factor $I/F$ (the ratio of reflected to incident flux) of Pluto's surface at our approach solar phase angle of 15° ranges from 0.1 in the dark equatorial regions to a peak of 0.7 in TR and the north polar cap. This is a wider range than any other solar system body except Iapetus (15).

Color imaging of the encounter hemisphere through three broadband filters (400 to 550 nm, 540 to 700 nm, and 780 to 975 nm) at 5 and 28 km/pixel spatial resolution reveals spectacular diversity across Pluto (Fig. 3). The bright, heart-shaped TR region divides into two distinct color units: The eastern half is more rugged, apparently physically thinner, and less red across the three broadband filters; this material may originate via some transport mechanism from SP. Dark equatorial regions (e.g., CR and Krun Macula) are particularly red at visible wavelengths and border a brighter region (exemplified by Viking Terra) to the north. At higher latitudes, this terrain grades into a unit that is bluer across the same three filters. We find that this unit brightens noticeably for high Sun elevations, a photometric behavior that contrasts with the flatter center-to-limb profiles of other Pluto regions and is potentially related to sea-sonal volatile ice sublimation. Interspersed with this bluer unit, especially above 60°N latitude, a redder unit appears. Contacts between these two high-latitude color units do not appear to consistently correlate with the underlying geomorphology and may be related to volatile transport processes.

Colors on Pluto are characteristic of refractory organic residues called tholins, which are readily formed by UV or charged-particle irradiation of mixtures of nitrogen and methane in both the gaseous and frozen states (16). Energetic radiation falling on Pluto's atmosphere and surface, each rich in nitrogen and methane, likely creates tholins that even in small concentrations yield colors ranging from yellow to dark red.

Ralph instrument images at a few IR wavelengths (e.g., fig. S2) have been downlinked to date. In Fig. 4, we show such images with 9 km/pixel spatial resolution, representing vibrational absorptions of $CH_4$ ice at 1.66 and 1.79 μm and CO ice at 1.58 μm. CO absorption, previously reported in ground-based Pluto spectra (17–19), is found to be strongest in SP's center (Fig. 4C). $CH_4$ ice is distributed widely, but the absorption depths vary from strong in the northern polar cap and SP to weak or nonexistent in the dark terrains (Fig. 4B). Sharp contrasts in $CH_4$ absorption correlate with geological units along the western edge of SP, with much weaker absorption associated with the bounding mountainous terrain.

The composition, extent, and uniqueness of SP's ices suggest that this region may be a major reservoir of volatile ices. It could also conceivably be a source region connected to the deep interior, or it could be a major sink for volatiles released planetwide, or both. Whether deep or surficial processes dominate is currently unclear, but the actual processes involved must be

**Fig. 3. Pluto color/ panchromatic composite image.** This is a composite of high-resolution pan-chromatic images and lower-resolution color images enhanced to show the diversity of surface units on Pluto; it was constructed from blue (400 to 550 nm), red (540 to 700 nm), and near-IR filter (780 to 975 nm) images from the Ralph instrument. The pan-chromatic observations were taken by the LORRI instrument from a distance of ~450,000 km from Pluto at a pixel scale of 2.2 km/pixel; the color observations were taken from a distance of

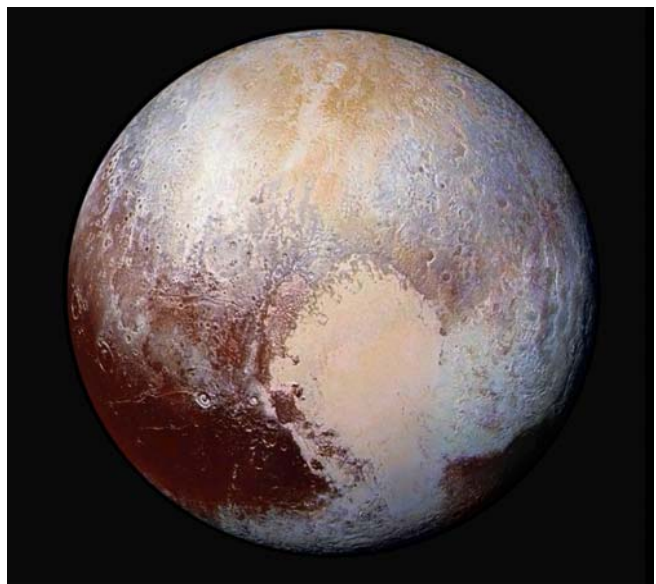

~250,000 km from Pluto at a pixel scale of 5.0 km/pixel.

**Fig. 4. Pluto Ralph surface imaging over the western side of Sputnik Planum.**
(**A**) Orthographic projection of the LORRI mosaic. (**B**) False-color Ralph image from three near-IR wavelengths (blue, 1.66 μm; green, 1.79 μm; red, 1.89 μm) selected to highlight methane ice absorption. Each color is mapped linearly from zero to the maximum reflectance at each wavelength. Regions with greater $CH_4$ absorption appear red; regions with weaker $CH_4$

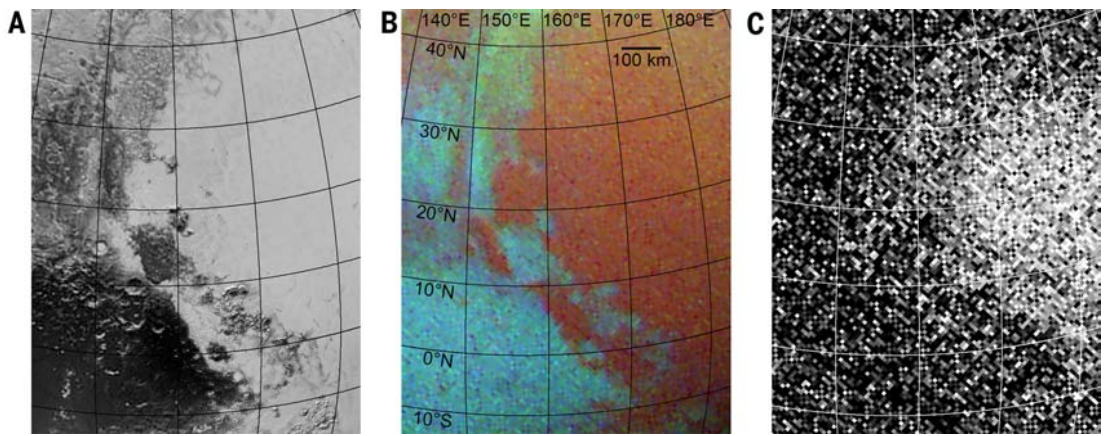

absorption appear blue-green. Regions with greater contrast between the 1.66-μm and 1.79-μm $CH_4$ absorptions tend toward yellow shades. Comparison with the associated LORRI mosaic shows a sharp transition from the strong $CH_4$ absorption on SP to much lower levels of $CH_4$ absorption on the montes along the west flank of SP as well as the portion of dark CR in the southwest corner. (**C**) LEISA (Linear Etalon Imaging Spectral Array) map of CO ice absorption in SP produced by subtracting the 1.58-μm image from the mean of the 1.57- and 1.59-μm images. Note the elevated CO abundance in the central region of SP indicated by the brighter pixels.

consistent with resupply of atmospheric $N_2$ and other volatiles against Pluto's rather prodigious atmospheric escape (20).

## Atmosphere

REX radio occultation measurements by New Horizons unequivocally reached Pluto's surface, providing the first direct measure of the temperature and pressure structure of the lower atmosphere (5). Preliminary results indicate that the surface pressure is ~10 μbar (5); this is lower than expected from the downward extrapolation of Earth-based stellar occultation measurements (21–24). At present it is unclear whether this reflects a recent decrease in the mass of the atmosphere—a reversal of the trend inferred from stellar occultations—or uncertainty in the relative calibration of the two techniques. These radio occultation measurements by New Horizons also suggest the presence of a shallow tropospheric boundary layer, consistent with recent predictions (25). High-altitude radio occultation data have not yet been sent to Earth but should provide ionospheric detections or constraints in the future.

High–phase angle images of Pluto made during flyby departure reveal a global atmospheric haze extending to ~150 km above the surface (Fig. 5A), with a derived normal optical depth of ~0.004. The high extent of the haze layer suggests a formation mechanism involving ion-molecule reactions or meteoritic dust. The atmospheric haze also shows structure, including possible waves and/or layering near 50 and 80 km altitude, which could be connected to buoyancy waves, as previously inferred from ground-based stellar occultation data (26).

UV solar occultation count rates have been sent to Earth; UV spectra themselves have not yet been. The occultation count rate data (Fig. 5B) show structure indicating absorption by $N_2$ starting at ~1670 km altitude, by $CH_4$ below ~960 km, by $C_2H_x$ hydrocarbons below 420 km, and haze below ~150 km. Ingress and egress observations made at opposite longitudes show nearly symmetric line-of-sight vertical absorption profiles (Fig. 5B), indicating a globally uniform upper atmospheric structure. These data are best fit with a $CH_4$ fractional number density abundance of ~0.25% (27), somewhat less than previous best estimates of 0.44% (28), indicating a slightly cooler atmosphere than expected. UV observations also indicate the discovery of two new atmospheric species from their far-UV absorption signatures: $C_2H_2$ and $C_2H_4$ at lower atmospheric mixing ratios of ~$3 \times 10^{-6}$ and ~$1 \times 10^{-6}$, respectively. Their opacities (and the solar occultation count rates) are consistent with a relatively stagnant atmosphere at 50 to 300 km altitude.

## Charon
### Geology and imaging

Our derived radius of Charon is $606 \pm 3$ km, similar to ground-based measurements (29); we also determined that Charon is not detectably oblate, with an upper bound on polar flattening of 1% (5). Substantial vertical relief of greater than 3 km is seen on the limb of Charon (fig. S3), which suggests that the widespread water ice seen spectroscopically across Charon is not a surface veneer and runs deep.

Charon mapping data that have arrived on Earth (Fig. 6) primarily cover the northern hemisphere and ranges from 32 km/pixel on the anti-Pluto (far approach) hemisphere to 4 km/pixel on the sub-Pluto (close approach) hemisphere (Fig. 6A; see also Fig. 2B). The only two images at ~400 m/pixel received to date reveal a complex geology characterized by numerous bright and dark spots, abundant fault scarps and darker curvilinear markings, both cratered and smooth plains, an extensive system of faults and graben, and a broad and prominent dark area centered on the north pole.

The dark polar spot, called Mordor Macula (Fig. 6B), is the most prominent albedo marking seen on Charon. This quasi-circular feature has a dark inner zone ~275 km across and roughly half as bright as the average surface of Charon (Fig. 1F). Its less dark outer zone is ~450 km across and fades gradually onto higher-albedo cratered plains. The inner zone of the dark spot is partly defined by a curvilinear marking that may be either a ridge or an exposed fault, indicating that this feature may be due to a large impact or complex tectonic structure, and suggests the possibility of a compositionally heterogeneous substrate.

Charon appears variably cratered across its surface, indicating variations in crater retention age. Both bright-rayed and dark-ejecta craters are also apparent at higher resolution (figs. S3 and S4). Such albedo variations may imply a compositionally variable surface, age effects, and/or impactor contamination.

A network of northeast-southwest–trending fractures cuts across most of the sub-Pluto hemisphere. The largest of these, called Macross and Serenity Chasmata (fig. S3), form a belt that extends at least 1050 km across the surface. Serenity Chasma is resolved as a double-walled graben-like structure, 60 km across at its widest and a few kilometers deep (Fig. 6D, 1). A deep trough observed on the limb at 30°N, 80°E has a depth of ~5 km. We interpret several dark curvilinear markings, observed on the less well-resolved anti-Pluto hemisphere, as global extensions of this fracture network.

An extensive area of rolling plains occurs south of the equator on Charon's sub-Pluto hemisphere (Fig. 6D, 2). The known extent of the plains, which stretch southward into the unimaged portions of Charon, is at least 400 km × 1000 km. These plains are moderately cratered and show several narrow rille-like features several kilometers wide when observed at 400 m/pixel (Fig. 6D, 3). Several large peaks of unknown origin extend 2 to 4 km above the rolling plains and are surrounded by moat-like depressions 1 to 3 km deep. The

**Fig. 5. Pluto LORRI and Alice atmospheric data.** (**A**) LORRI image of haze particle scattering in Pluto's atmosphere with solar phase angle of 167°. The haze exhibits a maximum $I/F$ of ~0.22 and extends to ~150 km altitude with a vertical scale height of 45 to 55 km. Its strong forward scattering suggests particles of ~0.5 μm effective diameter. (**B**) Total UV solar occultation count rates versus time. Horizontal scale is the time from center point of occultation. Black line shows ingress count rate; red (egress) count rate is overplotted in the reverse time direction to demonstrate their symmetry. The Sun's tangent altitude changes at 3.57 km/s during ingress and egress; the change in observed count rate is consistent with absorption by $N_2$ detected at ±800 s (~1670 km),

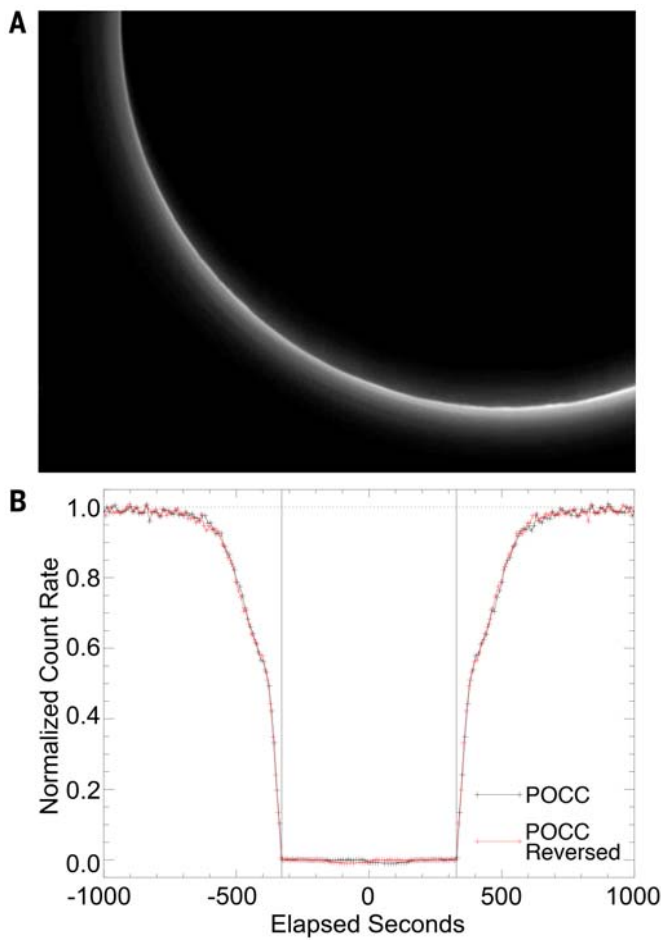

CH$_4$ at ±600 s (~960 km), higher hydrocarbons at ±450 s (~420 km), and possibly haze at ±375 s (~150 km).

most prominent of these, Kubrick Mons, is 20 × 25 km across and 3 to 4 km high (Fig. 6D, 4). Photoclinometry algorithms were used to estimate the relative elevations of these features, and they are consistent with shadow height measurements.

Craters were identified with some confidence on Vulcan Planum (Fig. 6 and fig. S5) because of the low Sun angles near the terminator and because of the generally level elevation of the terrain. For crater diameters of ≥10 km, we judge the cumulative areal crater density, $3 \times 10^{-4}$ to $4 \times 10^{-4}$ km$^{-2}$, to be reliable (fig. S5). Model ages can be assigned according to estimates of the impacting Kuiper Belt object (KBO) population (*11*, *12*). The KBO population is estimated at large (diameter ≥100 km) sizes from astronomical observations and can be extrapolated to smaller impactor sizes (the sizes that make the observable craters) under a variety of plausible assumptions; numerical integrations also provide estimates of the time rate of decay of the various Kuiper Belt subpopulations (*11*). These then provide a range of model ages for a given terrain with crater counts. For Vulcan Planum, most model ages from Greenstreet *et al.* (*11*) are ~4 billion years or older (i.e., equivalent to the presumed high-impact time period of the Late Heavy Bombardment or Nice model rearrangement of the outer solar system). Only the model age based on an estimate of the small KBO population from putative stellar occultations (*30*) (which has the largest number of small KBOs and thus the highest cratering rate) indicates that this region could be younger, perhaps 100 to 300 million years old.

## Surface brightness, color, and composition

Charon's panchromatic surface $I/F$ at our approach solar phase angle of 15° and available resolution ranges between 0.2 and 0.5, much more limited than Pluto's. Charon's north polar region is distinctly red at Ralph/MVIC (Multispectral Visible Imaging Camera) wavelengths, as shown in Fig. 6E. The reddish area encompasses the darkest region of the polar dark feature Mordor Macula but also extends well beyond. The boundary is indistinct and shows little correlation with geologic features.

One hypothesis for the reddish coloration is seasonal cold trapping of volatiles at Charon's poles followed by energetic radiation processing into more chemically complex, less volatile tholins that can remain after the pole emerges back into sunlight. Another possibility is a different composition at depth, as noted above.

### Atmosphere

As for Pluto, only solar occultation count rate data have arrived on Earth; no actual spectra have been downlinked as yet. The solar occultation total count rate showed sharp cutoffs at Charon ingress and egress, consistent with no atmosphere or an atmosphere far lower in column abundances than Pluto's. Upper limits were obtained for the vertical column densities of $N_2$ (~9 × 10$^{16}$ cm$^{-2}$), CH$_4$ (~5.6 × 10$^{15}$ cm$^{-2}$), and higher hydrocarbons (~2.6 × 10$^{15}$ cm$^{-2}$); much better constraints (or detections) will be possible when the solar occultation spectra are downlinked. No evidence of haze above Charon's limb is seen in high–phase angle (166°) imaging.

### Small satellites

Observations by New Horizons have provided the first spatially resolved measurements of Pluto's small moons Nix and Hydra; measurements of Styx and Kerberos have not yet been downlinked. We summarize these and other available results for Nix and Hydra next, and then report on our satellite and ring searches.

### Nix

A color composite image (Fig. 7A) shows a highly elongated body with dimensions of 49 × 32 km and an effective projected two-dimensional (2D) diameter of ~40 ± 2 km; a LORRI panchromatic image taken 128 s earlier gives essentially the same result. Another LORRI Nix image taken 8.73 hours earlier shows a nearly circular cross section with a projected 2D diameter of 34.8 ± 1 km. A triaxial ellipsoid shape with dimensions 54 × 41 × 36 km is consistent with both the resolved images and an extensive series of unresolved light curve measurements taken during the approach to Pluto, but Nix's mass is not yet well enough constrained to derive a reliable density. Nix shows evidence of compositional diversity in the color image, including a nonuniform distribution of red material possibly associated with a crater. Adopting reasonable phase laws of 0.02 to 0.03 mag/degree, we estimate that in visible light, the mean observed geometric albedo is 0.43 to 0.50. These high albedos indicate that Nix is likely covered with cleaner water ice than that on Charon.

### Hydra

Resolved panchromatic (but not color) measurements of Hydra are available (Fig. 7B); these show a highly nonspherical body with dimensions of roughly 43 km × 33 km (i.e., axial ratio of ~1.3) and an effective projected 2D diameter of ~41.1 ± 1 km. Surface albedo variations are seen, as are several crater-like features. Neither Hydra's mass nor its volume are well enough measured as yet to derive a reliable density. Hydra's average geometric albedo is 0.51 for a linear phase law coefficient of 0.02 mag/degree, derived from the observed brightness differences at the two epochs. Like Nix, Hydra has a highly reflective surface, which suggests relatively clean water ice. How such bright surfaces can be maintained on Nix and Hydra over billions of years is puzzling, given that a variety of external processes (e.g., radiation darkening, transfer of darker material from Charon via impacts, impacts with dark Kuiper Belt meteorites, etc.) would each tend to darken and redden the surfaces of these satellites over time.

**Fig. 6. Charon surface imaging results. (A)** Simple cylindrical mosaic of Charon. The area shown in detail in (D) is highlighted by a red boundary. **(B)** Polar stereographic mosaic of Charon's north pole. **(C)** Polar stereographic projection of Charon imagery as in (B). Contour lines are based on a color ratio between red and near-IR filters (0.54 to 0.70 μm and 0.78 to 0.98 μm, respectively) and highlight the two reddish color units discussed in the text (red, inner zone; yellow, outer zone). The red coloration is most pronounced in Mordor Macula but extends well beyond that central dark region. **(D)** Detail of Charon's sub-Pluto equatorial region [refer to red-outlined box in (A) for scale]; 400 m/pixel coverage can be seen at top left and bottom right. (1) Serenity Chasma, a graben ~60 km wide. (2) Sparsely cratered plains. (3) Surface fractures. (4) Kubrick Mons and its surrounding moat. **(E)** Enhanced color composite of Charon at 5 km/pixel resolution produced by placing blue, red, and near-IR Ralph instrument images (see Fig. 3) into the blue, green, and red color channels, and linearly stretching each color from zero.

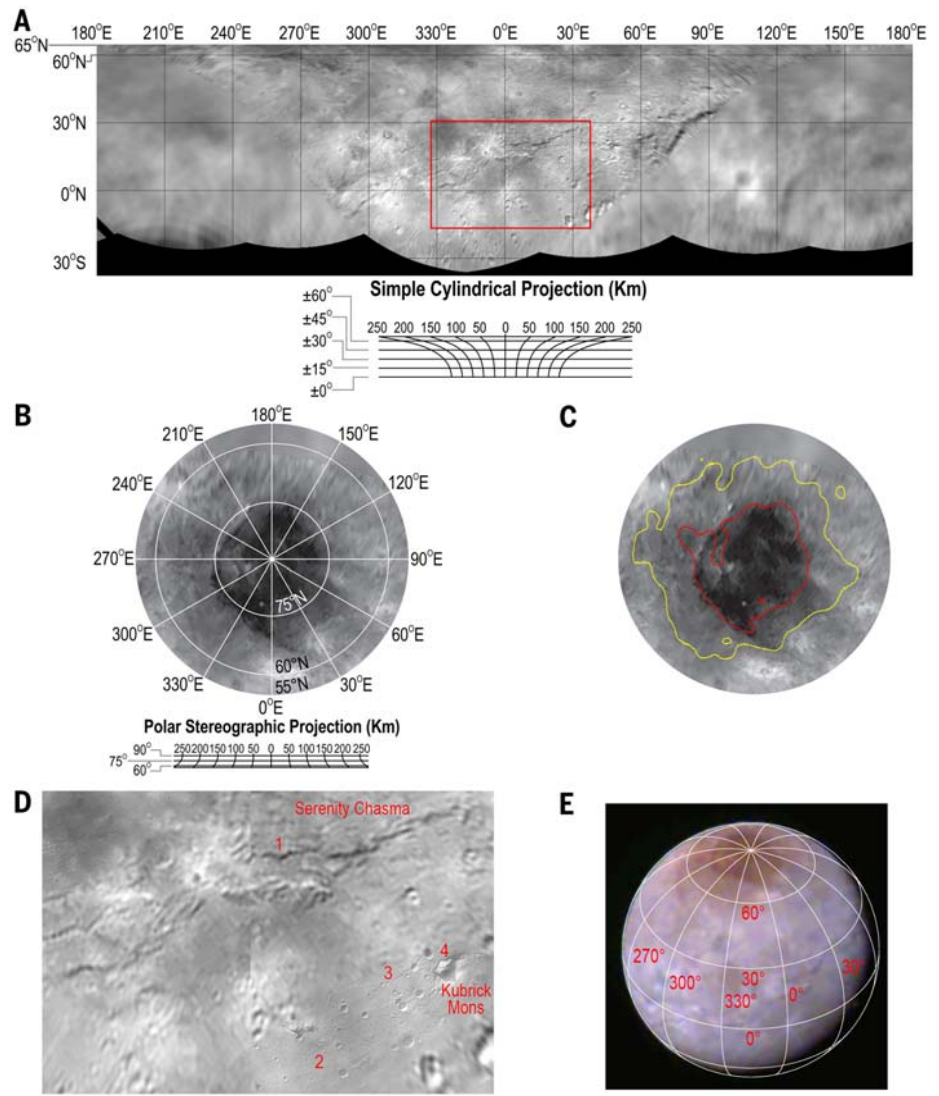

### Searches for small satellites and rings

New Horizons conducted seven deep searches for satellites and rings between 64 and 13 days before closest approach. No detections were made. For a Charon-like albedo of 0.38, diameter upper limits for undetected moons, determined by recovering model test objects implanted in the images, were 4.5 km at 110,000 to 180,000 km from Pluto, 2.4 km at 50,000 to 110,000 km from Pluto, 1.5 km at 19,000 to 50,000 km from Pluto (Charon is 19,600 km from Pluto), and 2.0 km at 5000 to 19,000 km from Pluto. No rings were found at an $I/F$ upper limit of $1.0 \times 10^{-7}$. These satellite and ring upper limits constitute substantial improvements over previous limits (*5, 31, 32*) (fig. S6).

**Fig. 7. LORRI and Ralph surface imaging of Nix and Hydra. (A)** Nix Ralph color composite image at 3.14 km/pixel with a solar phase angle of 15°, created by combining images taken through three filters (near-IR, 780 to 975 nm; red, 540 to 700 nm; blue, 400 to 550 nm) that were respectively loaded in the red, green, and blue planes of the color composite. **(B)** LORRI panchromatic (350 to 850 nm) image of Hydra at 1.13 km/pixel, solar phase angle 34°.

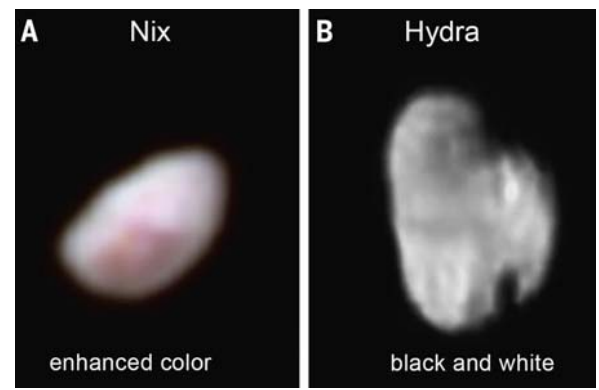

### Implications for origin and evolution

The New Horizons encounter with the Pluto system revealed a wide variety of geological activity—broadly taken to include glaciological and surface-atmosphere interactions as well as impact, tectonic, cryovolcanic, and mass-wasting processes (*13*)—on both the planet and its large satellite Charon. This suggests that other small planets of the Kuiper Belt, such as Eris, Makemake, and Haumea, could also express similarly complex histories that rival those of terrestrial planets such as Mars, as Pluto does.

It is notable that Triton—likely a Kuiper Belt planet captured by Neptune—was considered the best analog for Pluto before the encounter (13). However, our assessment is that the geologies of both worlds are more different than similar, although more quantitative consideration of this must await further data downlinks.

For Pluto, the rugged mountains and undulating terrain in and around TR require geological processes to have deformed and disrupted Pluto's water ice–rich bedrock. Some of the processes operating on Pluto appear to have operated geologically recently, including those that involve the water ice–rich bedrock as well as the more volatile, and presumably more easily mobilized, ices of SP and elsewhere. This raises questions of how such processes were powered so long after the formation of the Pluto system.

The bulk densities of Pluto and Charon differ by less than 10%, which is consistent with bulk rock contents for the two bodies that are likewise similar. Comparing models for the formation of the system by giant impact (7, 13) indicates that this similarity could imply that both precursor bodies were undifferentiated or only modestly differentiated before the collision, which would have profound implications for the timing, duration (33), and even the mechanism (34) of accretion in the ancestral Kuiper Belt.

### REFERENCES AND NOTES

1. S. A. Stern, The Pluto-Charon system. *Annu. Rev. Astron. Astrophys.* **30**, 185–233 (1992). doi: 10.1146/annurev.aa.30.090192.001153
2. J. K. Davies, J. McFarland, M. E. Bailey, B. G. Marsden, W.-H. Ip, in *The Solar System Beyond Neptune*, M. A. Barucci, H. Boenhardt, D. P. Cruikshank, A. Morbidelli, Eds. (Univ. of Arizona Press, Tucson, AZ, 2008), pp. 11–23.
3. S. A. Stern, The New Horizons Pluto Kuiper Belt mission: Overview with historical context. *Space Sci. Rev.* **140**, 3–22 (2008). doi: 10.1007/s11214-007-9295-y
4. H. A. Weaver, V. C. Gibson, M. B. Tapley, L. A. Young, S. A. Stern, Overview of the New Horizons science payload. *Space Sci. Rev.* **140**, 75–92 (2008). doi: 10.1007/s11214-008-9376-6
5. See supplementary materials on Science Online.
6. M. Person et al., The June 23 stellar occultation by Pluto: Airborne and ground observation. *Astron. J.* **146**, 83–98 (2013). doi: 10.1088/0004-6256/146/4/83
7. W. B. McKinnon, On the origin of the Pluto-Charon binary. *Astrophys. J.* **344**, L41–L44 (1989). doi: 10.1086/185526
8. J. Eluszkiewicz, D. J. Stevenson, Rheology of solid methane and nitrogen: Application to Triton. *Geophys. Res. Lett.* **17**, 1753–1756 (1990). doi: 10.1029/GL017i010p01753
9. Y. Yamashita, M. Kato, M. Arakawa, Experimental study on the rheological properties of polycrystalline solid nitrogen and methane: Implications for tectonic processes on Triton. *Icarus* **207**, 972–977 (2010). doi: 10.1016/j.icarus.2009.11.032
10. P. J. Tackley, Self-consistent generation of tectonic plates in time-dependent three-dimensional mantle convection simulations 1. Pseudoplastic yielding. *Geochem. Geophys. Geosyst.* **1**, 1021 (2000). doi: 10.1029/2000GC000036
11. S. Greenstreet, B. Gladman, W. B. McKinnon, Impact and cratering rates onto Pluto. *Icarus* **258**, 267–288 (2015). doi: 10.1016/j.icarus.2015.05.026
12. E. B. Bierhaus, L. Dones, Craters and ejecta on Pluto and Charon: Anticipated results from the New Horizons flyby. *Icarus* **246**, 165–182 (2014). doi: 10.1016/j.icarus.2014.05.044
13. J. M. Moore et al., Geology before Pluto: Pre-encounter considerations. *Icarus* **246**, 65–81 (2015). doi: 10.1016/j.icarus.2014.04.028
14. S. J. Peale, Tidally induced volcanism. *Celestial Mech. Dyn. Astron.* **87**, 129–155 (2003). doi: 10.1023/A:1026187917994
15. B. J. Buratti, J. A. Mosher, The dark side of Iapetus: Additional evidence for an exogenous origin. *Icarus* **115**, 219–227 (1995). doi: 10.1006/icar.1995.1093
16. D. P. Cruikshank, H. Imanaka, C. M. Dalle Ore, Tholins as coloring agents on outer solar system bodies. *Adv. Space Res.* **36**, 178–183 (2005). doi: 10.1016/j.asr.2005.07.026
17. T. C. Owen et al., Surface ices and the atmospheric composition of Pluto. *Science* **261**, 745–748 (1993). doi: 10.1126/science.261.5122.745; pmid: 17757212
18. W. M. Grundy, C. B. Olkin, L. A. Young, M. W. Buie, E. F. Young, Near infrared spectral monitoring of Pluto's ices: Spatial distribution and secular evolution. *Icarus* **223**, 710–721 (2013). doi: 10.1016/j.icarus.2013.01.019
19. W. M. Grundy, C. B. Olkin, L. A. Young, B. J. Holler, Near infrared spectral monitoring of Pluto's ices II: Recent decline of CO and N₂ absorptions. *Icarus* **235**, 220–224 (2014). doi: 10.1016/j.icarus.2014.02.025
20. K. N. Singer, S. A. Stern, On the provenance of Pluto's nitrogen (N₂). *Astrophys. J.* **808**, L50–L55 (2015). doi: 10.1088/2041-8205/808/2/L50
21. E. Lellouch et al., Pluto's lower atmosphere structure and methane abundance from high-resolution spectroscopy and stellar occultations. *Astron. Astrophys.* **495**, L17–L21 (2009). doi: 10.1051/0004-6361/200911633
22. B. Sicardy et al., Large changes in Pluto's atmosphere as revealed by recent stellar occultations. *Nature* **424**, 168–170 (2003). doi: 10.1038/nature01766; pmid: 12853950
23. J. L. Elliot et al., Changes in Pluto's atmosphere: 1988-2006. *Astron. J.* **134**, 1–13 (2007). doi: 10.1086/517998
24. C. B. Olkin et al., Evidence that Pluto's atmosphere does not collapse from recent occultations including the 2013 May 4 event. *Icarus* **246**, 220–225 (2015). doi: 10.1016/j.icarus.2014.03.026
25. A. M. Zalucha, X. Zhu, A. A. S. Gulbis, D. F. Strobel, J. L. Elliot, An analysis of Pluto's troposphere using stellar occultation light curves and an atmospheric radiative conductive convective model. *Icarus* **214**, 685–700 (2011). doi: 10.1016/j.icarus.2011.05.015

26. M. J. Person et al., Waves in Pluto's atmosphere. *Astron. J.* **136**, 1510–1518 (2008). doi: 10.1088/0004-6256/136/4/1510
27. G. R. Gladstone, Y. L. Yung, M. L. Wong, Pluto atmosphere photochemical models. *Lunar Planet. Sci.* **XLVI**, abstract 3008 (2015).
28. E. Lellouch et al., Exploring the spatial, temporal, and vertical distribution of methane in Pluto's atmosphere. *Icarus* **246**, 268–278 (2015). doi: 10.1016/j.icarus.2014.03.027
29. M. J. Person et al., Charon's radius and density from the combined data sets of the 2005 July 11 occultation. *Astron. J.* **132**, 1575–1580 (2006). doi: 10.1086/507330
30. H. E. Schlichting, C. I. Fuentes, D. E. Trilling, Initial planetesimal sizes and the size distribution of small Kuiper Belt objects. *Astron. J.* **146**, 36–42 (2013). doi: 10.1088/0004-6256/146/2/36
31. A. J. Steffl et al., New constraints on additional satellites in the Pluto system. *Astron. J.* **132**, 614–619 (2006). doi: 10.1086/505424
32. M. R. Showalter, D. P. Hamilton, Resonant interactions and chaotic rotation of Pluto's small moons. *Nature* **522**, 45–49 (2015). doi: 10.1038/nature14469; pmid: 26040889
33. W. B. McKinnon, D. Prialnik, S. A. Stern, A. Coradini, in *The Solar System Beyond Neptune*, M. A. Barucci, H. Boenhardt, D. P. Cruikshank, A. Morbidelli, Eds. (Univ. of Arizona Press, Tucson, AZ, 2008), pp. 213–241.
34. D. Nesvorný, A. N. Youdin, D. C. Richardson, Formation of Kuiper Belt binaries by gravitational collapse. *Astron. J.* **140**, 785–793 (2010). doi: 10.1088/0004-6256/140/3/785
35. M. Brozović, M. R. Showalter, R. A. Jacobson, M. W. Buie, The orbits and masses of satellites of Pluto. *Icarus* **246**, 317–329 (2015). doi: 10.1016/j.icarus.2014.03.015
36. G. L. Tyler et al., The New Horizons Radio Science Experiment (REX). *Space Sci. Rev.* **140**, 217–259 (2014). doi: 10.1007/s11214-007-9302-3

### ACKNOWLEDGMENTS

We thank M. Sykes and three anonymous referees for their careful work to improve this paper, and R. Tedford and C. Chavez for logistical support. We also thank the many engineers who have contributed to the success of the New Horizons mission and NASA's Deep Space Network (DSN) for a decade of excellent support to New Horizons. We acknowledge the contributions to New Horizons of our late colleagues David C. Slater and Thomas C. Coughlin. Supporting imagery is available in the supplementary material. As contractually agreed to with NASA, fully calibrated New Horizons Pluto system data will be released via the NASA Planetary Data System at https://pds.nasa.gov/ in a series of stages in 2016 and 2017 as the data set is fully downlinked and calibrated. This work was supported by NASA's New Horizons project.

### SUPPLEMENTARY MATERIALS

www.sciencemag.org/content/350/6258/aad1815/suppl/DC1
Materials and Methods
Supplementary Text
Figs. S1 to S6

5 August 2015; accepted 25 September 2015
10.1126/science.aad1815

**EMBARGOED UNTIL 2PM U.S. EASTERN TIME ON THE THURSDAY BEFORE THIS DATE:**